\documentclass[pra,twocolumn,a4paper,superscriptaddress]{revtex4}
\usepackage{graphicx,amsmath,bm,natbib, dsfont, longtable, amssymb}
\usepackage[bbgreekl]{mathbbol}
\usepackage[english]{babel}
\usepackage{graphicx,amsmath,bm, color}
\usepackage{latexsym}
\usepackage{amssymb}
\usepackage{amsfonts}
\usepackage{amsthm}
\usepackage{mathrsfs}
\usepackage{natbib}
\usepackage{ulem} 
\usepackage{hyperref}
\usepackage{cleveref}
\usepackage{verbatim,graphics, float}
\usepackage{psfrag}
\usepackage{subfigure}
\usepackage[usenames,dvipsnames,svgnames,table]{xcolor}
\providecommand{\openone}{\leavevmode\hbox{\small1\kern-3.8pt\normalsize1}} 

\providecommand{\bra}[1]{\langle#1|}
\providecommand{\ket}[1]{|#1\rangle}

\providecommand{\ketbra}[2]{|#1\rangle\kern-2.8pt\langle#2|}

\def\bra#1{\mathinner{\langle{#1}|}}
\def\ket#1{\mathinner{|{#1}\rangle}}

\def\text#1{\textrm{#1}}

\def\tr{\text{tr}}

\begin{document}

\title{Optimal entanglement witnesses in a split spin-squeezed Bose-Einstein condensate}
\date{\today}

\author{Enky Oudot}
\affiliation{Quantum Optics Theory, Department of Physics, University of Basel, Klingelbergstrasse 82, 4056 Basel, Switzerland}
\author{Jean-Daniel Bancal}
\affiliation{Quantum Optics Theory, Department of Physics, University of Basel, Klingelbergstrasse 82, 4056 Basel, Switzerland}
\author{Roman Schmied}
\affiliation{Quantum Atom Optics Lab, Department of Physics,University of Basel, Klingelbergstrasse 82, 4056 Basel, Switzerland}
\author{Philipp Treutlein}
\affiliation{Quantum Atom Optics Lab, Department of Physics,University of Basel, Klingelbergstrasse 82, 4056 Basel, Switzerland}
\author{Nicolas Sangouard}
\affiliation{Quantum Optics Theory, Department of Physics, University of Basel, Klingelbergstrasse 82, 4056 Basel, Switzerland}

\begin{abstract}
How to detect quantum correlations in bi-partite scenarios using a split many-body system and collective measurements on each party? We address this question by deriving entanglement witnesses using either only first or first and second order moments of local collective spin components. In both cases, we derive optimal witnesses for spatially split spin squeezed states in the presence of local white noise. We then compare the two optimal witnesses with respect to their resistance to various noise sources operating either at the preparation or at the detection level. We finally evaluate the statistics required to estimate the value of these witnesses when measuring a split spin-squeezed Bose-Einstein condensate. Our results can be seen as a step towards Bell tests with many-body systems. 
\end{abstract}
\maketitle


\section{Introduction} Substantial efforts have been devoted in the past years to the characterization of many-body systems through the entanglement of their elementary bodies \cite{Amico08, Bloch08}. While entanglement is usually detected using entanglement witnesses in many-body systems, first theoretical \cite{Mullin08, Laloe09, Gneiting08, Lewis_Swan15, Pelisson16, Tura14} and experimental \cite{Schmied16} steps have been taken to test a Bell inequality on a many-body system. 
The interest is twofold. First, the violation of a Bell inequality certifies the presence of a stronger form of quantum correlations than entanglement, namely Bell correlations \cite{Brunner14}. Second, Bell inequalities certify the presence of non-classical correlations device-independently, i.e. without assumption on the Hilbert space dimension or on the structure of the measurement operation \cite{Scarani12}. While Bell correlation witnesses have been proposed and used recently to successfully detect Bell-correlated states in a Bose-Einstein condensate \cite{Schmied16}, the device-independent detection of non-classical correlations remains to be demonstrated in many-body systems. The main problem is that Bell tests require to address the constituent bodies individually, which is challenging in many-body systems. A natural approach to circumvent this problem consists first in a bi-partite splitting of the constituent bodies and then, in applying collective measurements on each party. While the ultimate goal is to perform a Bell test, we focus on a simpler task in this manuscript, namely the detection of entanglement between these two parties. \\
 \\

\begin{figure} 
\includegraphics[width = 8.5cm]{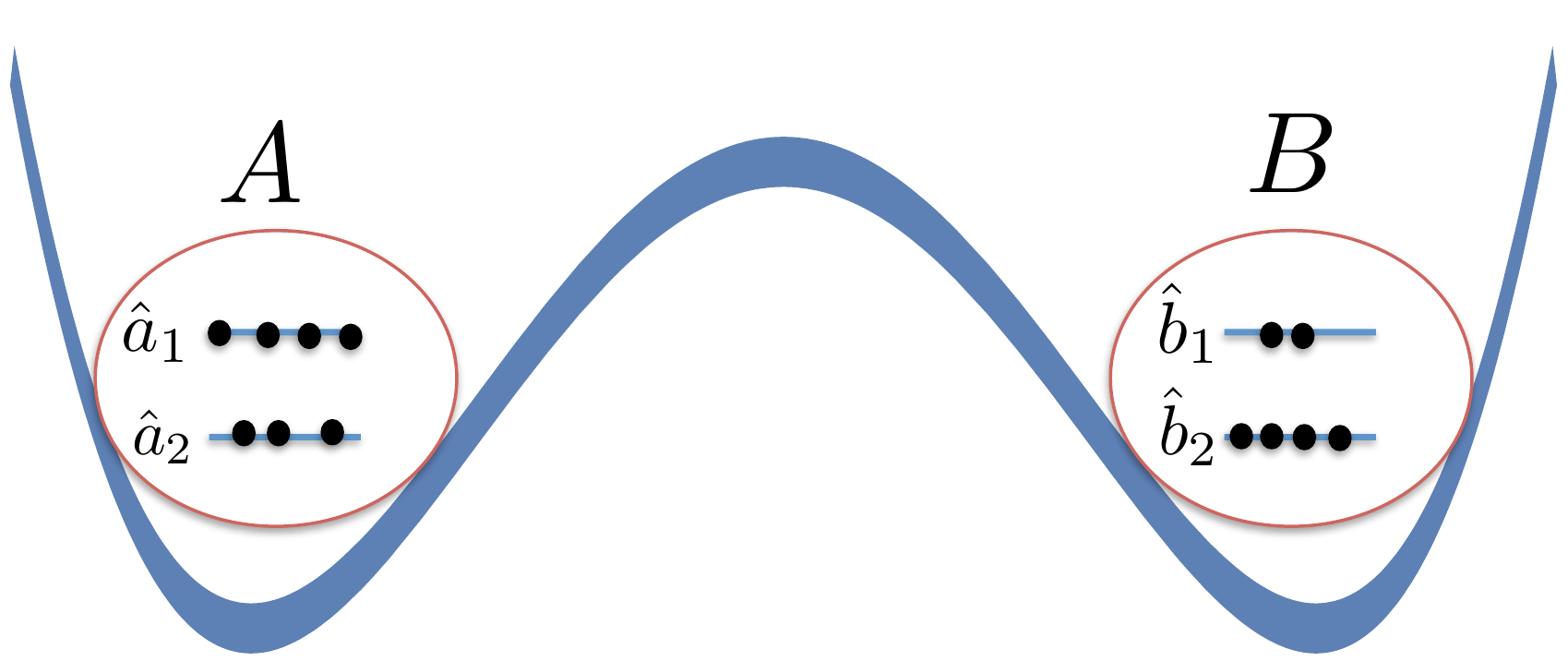}
\caption{Schematic representation of the four mode system of interest. The two internal states of the atoms located in A are initially prepared in a coherent spin state along the x direction \eqref{initial_state} before being squeezed with one-axis twisting \eqref{spinsqueezed_state}. The atoms are then spatially split and distributed between A and B with a binomial distribution before being measured collectively. The aim of this manuscript is to propose entanglement witnesses that could be used to reveal entanglement between the locations A and B in the presence of noise.}
\label{fig1}
\end{figure}

Let us clarify the scenario. We consider an ensemble of $N$ atoms with two internal states 1 and 2 and located at location A. Let $\hat{a}_i$ and $\hat{a}_i^\dag$ with $i\in \{1,2\},$ be the corresponding bosonic operators satisfying $[\hat{a}_i,\hat{a}_j^{\dagger}]=\delta_{i,j}$. To describe this ensemble of atoms, we use the picture of a collective spin, i.e. a vector of operators $\vec{J}^A$ with components 
\begin{eqnarray}
\label{collectivespin_xA}
 &&\hat{J}_x^A= \frac{1}{2}  (\hat{a}_1^{\dagger}\hat{a}_2+\hat{a}_1\hat{a}_2^{\dagger}),\\
\label{collectivespin_yA}
&&\hat{J}_y^A= \frac{1}{2 i} (\hat{a}_1^{\dagger}\hat{a}_2-\hat{a}_1\hat{a}_2^{\dagger}),\\
\label{collectivespin_zA}
&&  \hat{J}_z^A=\frac{1}{2}    (\hat{a}_1^{\dagger}\hat{a}_1-\hat{a}_2^{\dagger}\hat{a}_2)
\end{eqnarray}
satisfying the commutation relations 
\begin{equation}
[\hat{J}_i^A, \hat{J}_j^A]= i \epsilon_{ijk} \hat{J}_k^A
\end{equation}
where $\epsilon_{ijk}$ is the Levi-Civita symbol and $i,j,k \in \{x,y,z\}.$ The component $\hat{J}_z^A$ of the collective spin is half the population difference between the two internal states while  $\hat{J}_x^A$ and  $\hat{J}_y^A$ describe the coherence between these two states. We consider the case where initially this spin points in the x direction
\begin{equation}
\label{initial_state}
\ket{\psi_0}= \frac{1}{\sqrt{N}} e^{-i\frac{\pi}{2} \hat{J}_y^A} \hat{a}_{1}^{\dag  N} \ket{0}
\end{equation}
and then undergoes one-axis twisting \cite{Ma11, Pezze16}
\begin{equation}
\label{spinsqueezed_state}
\ket{\psi}=e^{-i\chi t (\hat{J}_z^A)^2} \ket{\psi_0}.
\end{equation}
This results in a spin-squeezed state, i.e. a state for which the variance along a certain direction $(\Delta \hat{J}_\bot^A)^2 = \langle  (\hat{J}_\bot^A) ^2 \rangle -\langle  \hat{J}_\bot^A \rangle^2 $ is smaller than $\frac{1}{N} | \langle \hat{J}_x^A \rangle|^2.$ This means that the mean spin projection of the state is large, and in a direction orthogonal to it, the spin variance is small. While the product of the squeezing rate $\chi$ and interaction time $t$ could be used to quantify the amount of squeezing as in Ref. \cite{Kitagawa93,Sorensen01}, one usually refers to the spin squeezing or Wineland parameter \cite{Wineland92, Wineland94} $\xi^2=\frac{N(\Delta \hat{J}_\bot^A)^2}{\langle \hat{J}_x^A\rangle^2}.$ For a coherent spin state like $\ket{\psi_0},$ $\xi^2=1.$  $\xi^2<1$ witnesses metrologically useful states, see e.g. \cite{Hammerer10, Pezze16} for a detailed discussion. For the state $\ket{\psi}$, this parameter is given by
\begin{eqnarray}
\nonumber
\xi^2=\frac{1}{4} \cos(\chi t )^{2-2N} \Big(3 + N -(N-1) \big(\cos(2 \chi t)^{N-2}  \\
\nonumber
+\sqrt{(1 - \cos(2 \chi t)^{N-2})^2 + 16 \cos(\chi t)^{2 N-4} \sin(\chi t)^2}\big)\Big).
\end{eqnarray}
In the rest of the paper, \textit{we quantify spin squeezing through the quantum noise reduction in dB using $10\log_{10}(\xi^2)$ for $N=500$ atoms. -10 dB squeezing for example corresponds to $\chi t =0.0058.$} Note that the existence of spin squeezing is connected to quantum correlations between the spins \cite{Kitagawa93} and many entanglement witnesses have been derived for spin squeezed states, see \cite{Ma11} and \cite{Pezze16} for reviews.\\

 In this manuscript, we consider the case where the atoms are spatially split with a state independent beamsplitter, i.e. 
\begin{equation}
\label{split_state}
\ket{\phi}=e^{\frac{\pi}{4}(\hat{a}_1^\dag \hat{b}_1 + \hat{a}_2^\dag \hat{b}_2 - \text{h.c.})} \ket{\psi}
\end{equation}
where $\hat b_i$ and $\hat b_i^{\dagger}$ are bosonic operators for the location B, see Fig. \ref{fig1}. Our aim is to show how to reveal entanglement between A and B using the collective spin observables given in Eqs. \eqref{collectivespin_xA}--\eqref{collectivespin_zA} and similarly for B. Let us mention that entanglement \cite{He12,Bar2011,He11,Kurkjian13} and steering \cite{Opanchuk12} have been studied in a different scenario where a beam splitter interaction is applied in order to couple two spin squeezed states. In this manuscript, we show how to derive optimal witnesses for the state $|\phi\rangle$ in the presence of local white noise using either only first or first and second order moments of local collective spins. Interestingly, we find in each case witnesses that are closely related to existing entanglement criteria \cite{Duan00,Simon00, Raymer03, Durkin05} and we show how they could be used to reveal entanglement in a split Bose-Einstein condensate (BEC).\\

Concretely, we consider a two-component BEC of alkali atoms where two hyperfine states represent a pseudo-spin $\frac{1}{2}$ for each atom, see Fig. \ref{fig1}. Such a BEC can be prepared in one of the two hyperfine levels without discernible thermal component before being rotated with a $\pi/2$ pulse around the $y$ axis, hence creating a coherent spin state pointing along the x-direction as described by \eqref{initial_state}. To create quantum correlations between the spins, one can make use of elastic collisions in state dependent potentials \cite{Riedel2010,Gross2010}, giving rise to one-axis twisting as in Eq. \eqref{spinsqueezed_state}. The spatial splitting is done by slowly raising a barrier in a state-independent potential as in Refs. \cite{Shin04, Schumm05}. To characterize the resulting state, the collective observables $\hat{J}_z^{A/B}$ can be accessed locally in each well by counting the numbers of atoms in each hyperfine state using resonant absorption imaging \cite{Reinaudi2007}. Projections along other spin directions are obtained by appropriate Rabi rotations in each well before the measurement. We show through a detailed feasibility study that the detection of entanglement in this system in within reach using currently available setups.\\

The outline of this paper is the following. In section \ref{SectionII}, we derive witnesses using first order moments of local collective spin operators, i.e.  $\langle\hat{J}_{i}^A\rangle$, $\langle\hat{J}_{i}^B\rangle$,  $\langle\hat{J}_{i}^A \hat{J}_{j}^B\rangle$ where $i, j$ labels the components in the directions $x$, $y$ and $z.$  We show in particular, the entanglement witness that is optimal regarding the tolerance to local white noise. In section \ref{SectionIII}, we consider the set of witnesses involving not only first order moments of local collective operators, but also the second order moments $\langle(\hat{J}_{i}^A)^2\rangle$ and $\langle(\hat{J}_{i}^B)^2\rangle$ and derive again the witness that is optimal with respect to the tolerance to local white noise. The optimal witnesses presented in sections \ref{SectionII} and \ref{SectionIII} are then compared in section \ref{SectionIV} with respect to various experimental issues operating either at the level of the state preparation or at the level of the detection. The section \ref{SectionV} is devoted to a feasibility study using a spin-squeezed Bose-Einstein condensate. We quantify in particular the statistics that is needed to estimate the value of our entanglement witnesses in realistic parameter regimes. We conclude in the last section.


\section{Entanglement witnesses using first order moments of local collective spin observables}
\label{SectionII}
This section is divided into three subsections. The first one shows how to derive entanglement witnesses using first order moments of local collective spin observables. The second subsection aims at identifying the witness that is optimal with respect to local white noise. The last subsection presents the result of this optimization.
\subsection{Construction of entanglement witnesses}
We first consider the case where $n_a$ atoms are located in A and $n_b$ in B. With this in mind, we focus on the set of expectation values of first order moments of local collective spin observables (LCSO). This is a real space consisting of all possible values of $\langle\hat{J}_i^{A}\rangle $, $ \langle\hat{J}_i^{B}\rangle $, $\langle\hat{J}_{i}^{A} \hat{J}_{j}^{B}\rangle$ where $i,j,k=\{x,y,z\}.$ Note that the marginals $\langle\hat{J}_i^{A}\rangle $ and $ \langle\hat{J}_i^{B}\rangle $ are constrained by
\begin{equation} 
\label{normeJ}
 ||\langle\vec{J}^A\rangle||\leq \frac{n_a}{2}  ~~~~~~~~~  ||\langle\vec{J}^B\rangle||\leq \frac{n_b}{2} .
 \end{equation}
 This can be seen by noting that by a rotation, the vector $\langle\vec{J^A}\rangle=(\langle\hat{J}_x^A\rangle,\langle\hat{J}_y^A\rangle,\langle\hat{J}_z^A\rangle)$ can be brought to a form where one component only is non vanishing. As any component $\hat{J}_i^{A}$ has $-n_a/2$ and $n_a/2$ as eigenvalues with the largest modulus, $||\langle\vec{J}^A\rangle||$ is bounded by $n_a/2.$ The same arguments apply to $||\langle\vec{J}^B\rangle||.$ We call $\mathcal{U}$ the space of possible values of $\langle\hat{J}_i^{A}\rangle $, $ \langle\hat{J}_i^{B}\rangle $, $\langle\hat{J}_{i}^{A} \hat{J}_{j}^{B}\rangle$ satisfying the inequality \eqref{normeJ}.

\begin{figure}
\includegraphics[width = 7.5cm]{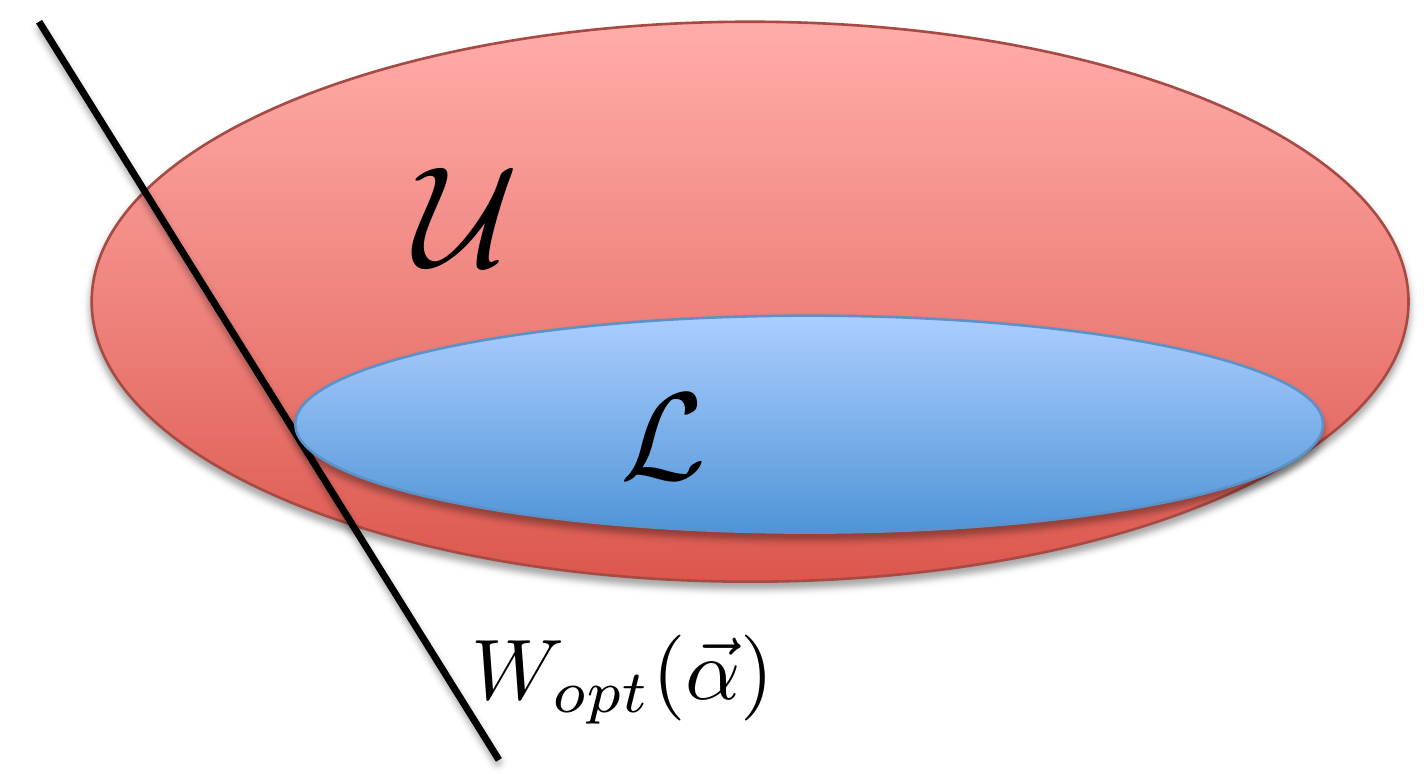}
\caption{Schematic representation of the set $\mathcal{U}$ of expectation values of first order moments of local collective spin observables $\langle\hat{J}_i^{A}\rangle $, $\langle\hat{J}_i^{B}\rangle $, $\langle\hat{J}_{i,j}^{A,B}\rangle$ where $i,j,k\in\{x,y,z\}$ satisfying $||\langle\vec{J}^A\rangle||\leq \frac{n_a}{2}$ and $||\langle\vec{J}^B\rangle||\leq \frac{n_b}{2}.$
The subset $\mathcal{L}$ is generated by the expectation values of first order moments of local collective spin observables obtained from separable states \eqref{sep}. Since $\mathcal{L}$ is a convex set, a family of linear witnesses is sufficient to fully characterize it.}
\label{fig2}
\end{figure}

We now consider a subspace $\mathcal{L}$ generated by the expectation values of first order moments of LCSO that are obtained from separable states, i.e. states of the form  
 \begin{equation}
 \label{sep}
 \rho_{n_a, n_b}=\sum_k p_k \rho^{A{(k)}}_{n_a}\otimes\rho^{B{(k)}}_{n_b}
 \end{equation} 
where $p_k$ is a probability distribution. $\mathcal{L}$ is a convex set. This can be seen by considering the sum $\Lambda \vec{X} + (1-\Lambda) \vec{Y}$ of two vectors in $\mathcal{L}$ where $\Lambda$ is an arbitrary positive real number smaller than or equal to $1.$ The components of $\Lambda \vec{X} + (1-\Lambda) \vec{Y}$ can be written as a sum of two traces involving the same LCSO and two different separable states. By the linearity of the trace and the convexity of the set of separable states, we deduce that $\Lambda \vec{X} + (1-\Lambda) \vec{Y}$ belongs to $\mathcal{L},$ i.e. $\mathcal{L}$ is convex. Hence, to characterize $\mathcal{L},$ it is sufficient to consider witnesses that are linear with respect to $\langle\hat{J}_i^{A}\rangle$, $\langle\hat{J}_i^{B}\rangle$, $\langle\hat{J}_{i}^{A} \hat{J}_{j}^{B}\rangle,$ see Fig. \ref{fig2}.
Such witnesses are of the form $W(\vec{\alpha})_{n_a,n_b} = \langle \hat W(\vec{\alpha}) \rangle$ with
\begin{equation}
 \label{witness_obs}
\hat{ W}(\vec{\alpha})=
   \sum\limits_{\substack{i,j=x,y,z}} \alpha_{i,j} \hat{J}_{i}^A \hat{J}_{j}^B
   + \bar{\alpha_{i}} \hat{J}_{i}^A +\alpha_{i} \hat{J}_{i}^B
\end{equation}
the corresponding operators. These witnesses can be parametrized by a vector $\vec{\alpha}=(\alpha_{i,j},\bar{\alpha_i},\alpha_i)$ with 15 elements. Each vector $\vec{\alpha}$ defines one particular direction in the space $\mathcal{U}$ and the maximum value that a given $W(\vec{\alpha})$ can take over the set of separable states defines the boundary of $\mathcal{L}$ in the direction $\vec{\alpha}$.

For any separable state of the form \eqref{sep}, we have  
\begin{eqnarray}
\label{sep_bound}
\nonumber
  W(\vec{\alpha})_{sep,n_a,n_b}\leq
   \max_k \Big(\sum\limits_{\substack{i,j=x,y,z}} \alpha_{i,j}\langle\hat{J}_{i}^{A(k)} \rangle\langle\hat{J}_{j}^{B(k)}\rangle \\ 
   + \bar{\alpha_{i}}\langle\hat{J}_{i}^{A(k)}\rangle +\alpha_{i}\langle\hat{J}_{i}^{B(k)}\rangle\Big)
  \end{eqnarray}
where $W(\vec{\alpha})_{sep,n_a,n_b}$ refers to the set of values attainable by $W(\vec{\alpha})_{n_a,n_b}$ while considering only the separable states given in Eq. \eqref{sep}, $\langle\hat{J}_{i}^{A(k)} \rangle=\tr \left( \rho^{A{(k)}}_{n_a} \hat{J}_{i}^A \right)$ and similarly for $\langle\hat{J}_{i}^{B(k)} \rangle.$ For a given choice of $\vec{\alpha}$, the value of $k$ which saturates the inequality \eqref{sep_bound} defines a separable bound $w(\vec{\alpha})_{n_a,n_b},$ i.e. the maximum value that $W(\vec{\alpha})_{sep,n_a,n_b}$ can take. The latter can be computed as
\begin{eqnarray}
\nonumber
 w(\vec{\alpha})_{n_a,n_b}=
   \max_{ ||\langle\vec{J}^A\rangle||\leq \frac{n_a}{2}, ||\langle\vec{J}^B\rangle||\leq \frac{n_b}{2}} \Big(\sum\limits_{\substack{i,j=x,y,z}} \alpha_{i,j}\langle\hat{J}_{i}^{A} \rangle\langle\hat{J}_{j}^{B}\rangle \\ 
   \nonumber
   + \bar{\alpha_{i}}\langle\hat{J}_{i}^{A}\rangle +\alpha_{i}\langle\hat{J}_{i}^{B}\rangle\Big).
  \end{eqnarray}
This yields the following family of witnesses
\begin{equation}
\label{witness}
w(\vec{\alpha})_{n_a,n_b} -W(\vec{\alpha})_{n_a,n_b} \geq 0
\end{equation}
which are satisfied by measurement on all separable states. A violation of this inequality reveals the presence of entanglement.\\

Now consider the case in which $N$ spins are split leading to a fluctuating number of particle between the two locations A and B at each run. Since we are only considering local spin observable measurements, the coherence between different atom numbers on each side cannot be probed and only the distribution of the particles $p(n_a,N-n_a)$ between the two wells matters. Following the same line of thought we get a separable bound for any distribution of particles across the two wells, including the case where the atomic fluctuations during the splitting result in reduced fluctuations of the relative atom number between A and B. That is, $w(\vec{\alpha})=\sum_{n_a} p(n_a,N-n_a) w(\vec{\alpha})_{n_a,N-n_a}.$ Since we are considering the splitting given in Eq.  \eqref{split_state} leading to a binomial distribution of particles, we end up with the separable bound 
\begin{equation}
w(\vec{\alpha})=\sum_{n_a}\frac{1}{2^N}\binom{N}{n_a}w(\vec{\alpha})_{n_a,N-n_a}
\end{equation}
and the corresponding entanglement witnesses
\begin{equation}
\label{witness_bino}
w(\vec{\alpha}) -W(\vec{\alpha}) \geq 0
\end{equation}
with $W(\vec{\alpha})$ the expectation value of $\hat{W}(\vec{\alpha})$ given in Eq. \eqref{witness_obs}, evaluated on the state \eqref{split_state} which involves variable local atom numbers.

\subsection{Optimal witness with respect to local white noise}
Now that a family of witnesses is available, we want to find the one that is the most relevant for the scenario described in the introduction. In particular, we consider the general case where the split spin squeezed state $\ket{\phi}$ experiences local white noise in each location, i.e. we consider the state
\begin{equation}
\label{noisy_state}
\rho_{\text{noisy}}=p\ket{\phi}\bra{\phi}+\sum\limits_{\substack{k=0}}^N \frac{(1-p)}{(k+1)(N-k+1)}\binom {N} {k}\mathbb{I}_k\otimes\mathbb{I}_{N-k}
\end{equation}
where $\mathbb{I}_{k}$ is the identity for $k$ particles in the symmetric subspace and we look for the witness that can detect entanglement for the smallest value of $p.$ Note first that 
$W(\vec{\alpha})_{\rho_\text{noisy}}=\tr(\hat{W}(\vec{\alpha})\rho_{\text{noisy}})= p~ \tr(\hat{W}(\vec{\alpha})\ket{\phi}\bra{\phi}) = p W(\vec{\alpha})_{\ket{\phi}}.$ For a given choice of $\vec\alpha,$ we define $W(\vec{\alpha})_{\ket{\phi}}^{\text{opt}}$ as the maximal value of $W(\vec{\alpha})_{\ket{\phi}}$ over all possible local rotations. Since entanglement is by definition invariant under local rotation, the resistance to noise of the witness corresponding to the direction $\vec{\alpha}$ is given by
\begin{equation}
p W(\vec{\alpha})_{\ket{\phi}}^{\text{opt}} = w(\vec{\alpha}).
\end{equation}  
The optimal witness is thus associated with the particular direction $\vec{\alpha}$ such that the ratio $w(\vec{\alpha})/W(\vec{\alpha})_{\ket{\phi}}^{\text{opt}}$ takes the smallest possible value. Since the state $\ket{\phi}$ depends on $\chi t$ and $N,$ the procedure needs to be repeated when changing these two parameters. The result of this optimization is given in the next subsection.

 \subsection{Result of the optimization}
\begin{figure}[H] 
\includegraphics[width = 8.5cm]{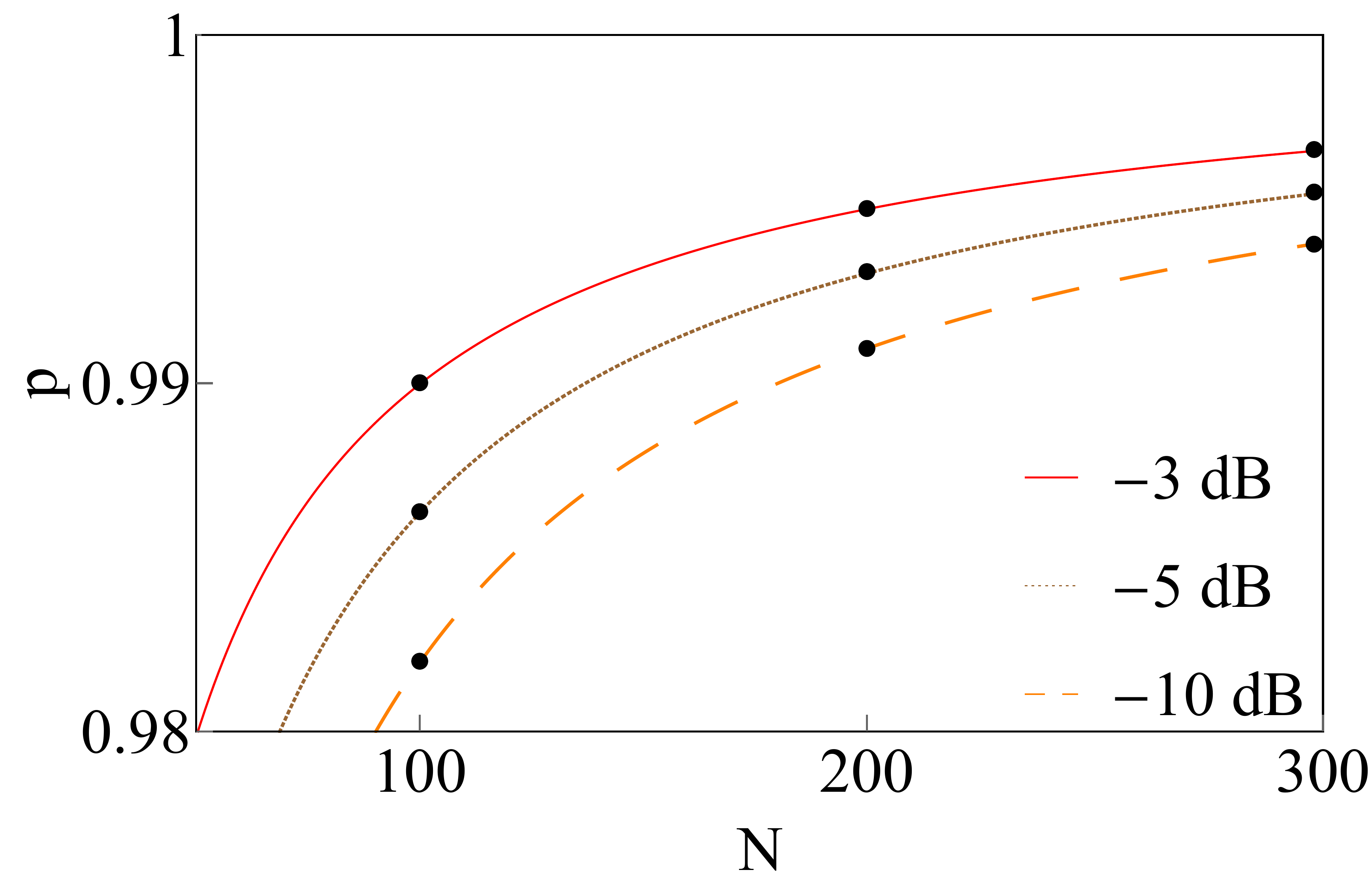}
\caption{Tolerable noise as a function of the number of atoms for the criterion \eqref{S_criterion} for different squeezing. The black dots correspond to the results of the numerical optimization, which have been obtained following the procedure described in B.}
\label{fig5}
\end{figure}
In order to find the witness admitting the largest amount of noise, we minimized numerically the value of the ratio $w(\vec{\alpha})/W(\vec{\alpha})_{\ket{\phi}}^{\text{opt}}$ over all choices of $\vec \alpha$ and of local unitaries. We display the results of this optimization (black dots) in Figure 3 where we plot the resistance of noise vs. the spin number for various squeezing parameters. For comparison, we also plot the resistance of the criterion S (solid, dashed and dotted lines) whose precise form is given below in a basis where the state $\ket{\phi}$ is rotated by the squeezing angle around the $x$ axis before the beamsplitter so that $z$ corresponds to the squeezed direction \cite{Pezze16}  
\begin{equation}
\label{S_criterion}
\text{S}=\langle\hat{J}_x^A\hat{J}_x^B\rangle+\langle\hat{J}_y^A\hat{J}_y^B\rangle-\langle\hat{J}_z^A\hat{J}_z^B\rangle\leq\frac{N(N-1)}{16}.
\end{equation}
The previous inequality holds for any separable state. It is closely connected to the minimization of the scalar product between $\vec{J}^A$ and $\vec{J}^B$ \cite{Durkin05}, which requires correlations between the two parties to be violated, namely entanglement. The comparison in Figure 3 shows that this scalar product is the optimal entanglement witness involving first order moments of LCSO for spin squeezed states with local white noise in the considered parameter region.

\section{Entanglement witnesses using second order moments of local collective spin observables}
\label{SectionIII}
In this section, we follow the line of thought presented in the previous section to develop entanglement witnesses involving higher order moments. We start by considering the real space  consisting of all possible values of $\langle\hat{J}_i^{A}\rangle $, $ \langle\hat{J}_i^{B}\rangle $, $\langle\hat{J}_{i}^{A}\hat{J}_{j}^{B}\rangle$, $\langle(\hat{J}_i^{A})^2\rangle$ and $\langle(\hat{J}_i^{B})^2\rangle$ satisfying the constraints    
\begin{eqnarray}
\label{cons1}
&||\langle\vec{J}^A\rangle||\leq \frac{n_a}{2} \\ 
\label{cons2}
& \langle(\hat{J}^A_x)^2\rangle+\langle (\hat{J}^A_y)^2 \rangle+\langle (\hat{J}^A_z)^2 \rangle\leq \frac{n_a}{2}\left(\frac{n_a}{2}+1\right)\\
\label{cons3}
& (\Delta \hat{J}_i^A)^2 =\langle(\hat{J}^A_i)^2\rangle-\langle\hat{J}^A_i \rangle^2\geq 0 \\
\label{cons4}
& (\Delta \hat{J}_i^A)^2 + (\Delta \hat{J}_j^A)^2- |\langle \hat{J}_k^A\rangle| \geq 0 \\ 
\nonumber
\end{eqnarray}
and similarly for $B$ and $i,j,k \in \{x,y,z\}.$ Note that we do not consider higher order moments like $\langle(\hat{J}_i^{A})^2\hat{J}_j^{B}\rangle$ as they often require more experimental runs to be evaluated. According to angular momentum theory, the second and the third constraints are valid for all quantum states, the fourth one comes from the Heisenberg inequality. Since the space of first and second order moments of LCSO is convex, we look again for witnesses that are linear in the parameters given above. Let us consider the quantity
\begin{eqnarray}
 \label{2ndorder}
 \nonumber
  W_{2}(\vec{\alpha})_{na,nb}=
   \sum\limits_{\substack{i,j=x,y,z}} \Big(\alpha_{i,j}\langle\hat{J}_{i}^A\hat{J}_{j}^B\rangle +\bar{\alpha}_{i} \langle\hat{J}_{i}^A\rangle + \alpha_{i} \langle\hat{J}_{i}^B\rangle
   \nonumber \\
   \nonumber +
   \bar{\alpha}_{i}^{(2)}\langle(\hat{J}_{i}^A)^2\rangle+\alpha_{i}^{(2)}\langle(\hat{J}_{i}^B)^2\rangle \Big). 
\end{eqnarray} 
$\vec{\alpha}$ is here a vector with 21 elements $(\alpha_{i,j},\bar{\alpha}_{i},{\alpha}_{i},\bar{\alpha}_{i}^{(2)},\alpha_{i}^{(2)}).$ When the expectation values are taken on the set on separable states, the previous quantity can be upper bounded by 
\begin{eqnarray}
 \label{2ndorder}
 \nonumber
  w_{2}(\vec{\alpha})_{na,nb}=\max_{\vec{J}^A, \vec{J}^B} 
   \sum\limits_{\substack{i,j=x,y,z}} \Big(\alpha_{i,j} \langle\hat{J}_{i}^A\rangle \langle \hat{J}_{j}^B \rangle +\bar{\alpha}_{i} \langle\hat{J}_{i}^A\rangle 
 \nonumber \\
   + \alpha_{i} \langle\hat{J}_{i}^B\rangle + \bar{\alpha}_{i}^{(2)}\langle(\hat{J}_{i}^A)^2\rangle+ \alpha_{i}^{(2)}\langle(\hat{J}_{i}^B)^2\rangle \Big),
\end{eqnarray} 
where the maximum is computed from the set of vectors $\vec{J}^A,$ $\vec{J}^B$ satisfying~\eqref{cons1} - \eqref{cons4}. This yields the following family of entanglement witnesses suited for spins distributed binomially between the locations $A$ and $B$
\begin{equation}
\label{witness_bino2}
w_{2}(\vec{\alpha}) - W_2(\vec{\alpha}) \geq 0
\end{equation}
where 
\begin{equation}
w_2(\vec{\alpha})=\sum_{n_a}\frac{1}{2^N}\binom{N}{n_a}w_2(\vec{\alpha})_{n_a,N-n_a}
\end{equation}
and $W_2(\vec{\alpha}) = \langle\hat{W}_2(\vec{\alpha})\rangle$ with
\begin{eqnarray}
 \label{2ndorder}
\hat{W}_{2}(\vec{\alpha})=
   \sum\limits_{\substack{i,j=x,y,z}}  \Big(\alpha_{i,j} \hat{J}_{i}^A\hat{J}_{j}^B+\bar{\alpha}_{i} \hat{J}_{i}^A + \alpha_{i} \hat{J}_{i}^B
    \\
    +\bar{\alpha}_{i}^{(2)} (\hat{J}_{i}^A)^2 + \alpha_{i}^{(2)} (\hat{J}_{i}^B)^2 \Big). 
\end{eqnarray}

Now consider states of the form \eqref{noisy_state}. As before, we optimize $W_2(\vec{\alpha})_{\ket{\phi}} = \tr\left(\hat{W}_{2}(\vec{\alpha}) |\phi\rangle\langle \phi| \right)$ over all possible local rotations for a given choice $\vec{\alpha}$. This defines $W_2(\vec{\alpha})_{\ket{\phi}}^{\text{opt}}.$ We then extract the minimum value of $p$ for each witness from the equation
\begin{equation}
pW_2(\vec{\alpha})_{\ket{\phi}}^{\text{opt}} + (1-p) \sum_{i=x,y,z} \frac{\alpha_{i}^{(2)}+\bar{\alpha}_{i}^{(2)}}{12} N(N+5) = w_{2}(\vec{\alpha}),
\end{equation}
where the second term in the left hand side comes from the mean values of second order moments of LCSO on local white noise. The optimal witness is then obtained by looking for the direction $\vec{\alpha}$ leading to the minimum value of $p.$ Note that this optimization is not particularly easy as it is a nonlinear optimization and the space of possible values of first and second order moments of LCSO has a dimension 21. To make it simpler, we restrict our interest to symmetric witnesses only -- note that the state on which we are optimizing is also symmetric under exchange of parties. Over 6000 numerical optimizations with $N=26$ atoms and a squeezing corresponding to $\chi t =0.0058$ before splitting, we found the following optimal witness twice
\begin{equation}
\label{Duan_criterion}
\text{D}=\langle (\hat{J}_y^A-\hat{J}_y^B)^2 \rangle + \langle (\hat{J}_z^A+\hat{J}_z^B)^2 \rangle - \langle \hat{J}_x^A+\hat{J}_x^B\rangle \geq 0.
\end{equation}
This witness is satisfied for all separable states. We have not been able to find a better witness for any value of $\chi t$ corresponding to squeezing parameters between $-1$ and $-10$ dB for 500 atoms and for any atom number between 25 and 100 atoms.

The witness \eqref{Duan_criterion} is again given in a basis where $\ket{\phi}$ is rotated by the squeezing angle around the $x$ axis before the beamsplitter so that $z$ corresponds to the squeezing direction. Note that this criterion can be seen as a linear form of the well known Duan \cite{Duan00} and Simon \cite{Simon00} criteria that have been successfully used for witnessing continuous variables entanglement more than 15 years ago \cite{Julsgaard01}, see also the generalization in Ref. \cite{Raymer03}. By linear, we mean that D involves the mean values $\langle\hat{J}_i^{A}\rangle $, $ \langle\hat{J}_i^{B}\rangle$, $\langle\hat{J}_{i}^{A}\hat{J}_{j}^{B}\rangle$, $\langle(\hat{J}_i^{A})^2\rangle$ and $\langle(\hat{J}_i^{B})^2\rangle$ only while the criteria  \cite{Duan00, Simon00, Raymer03} also use the square of these mean values.

\begin{figure}[H] 
\includegraphics[width = 8.5cm]{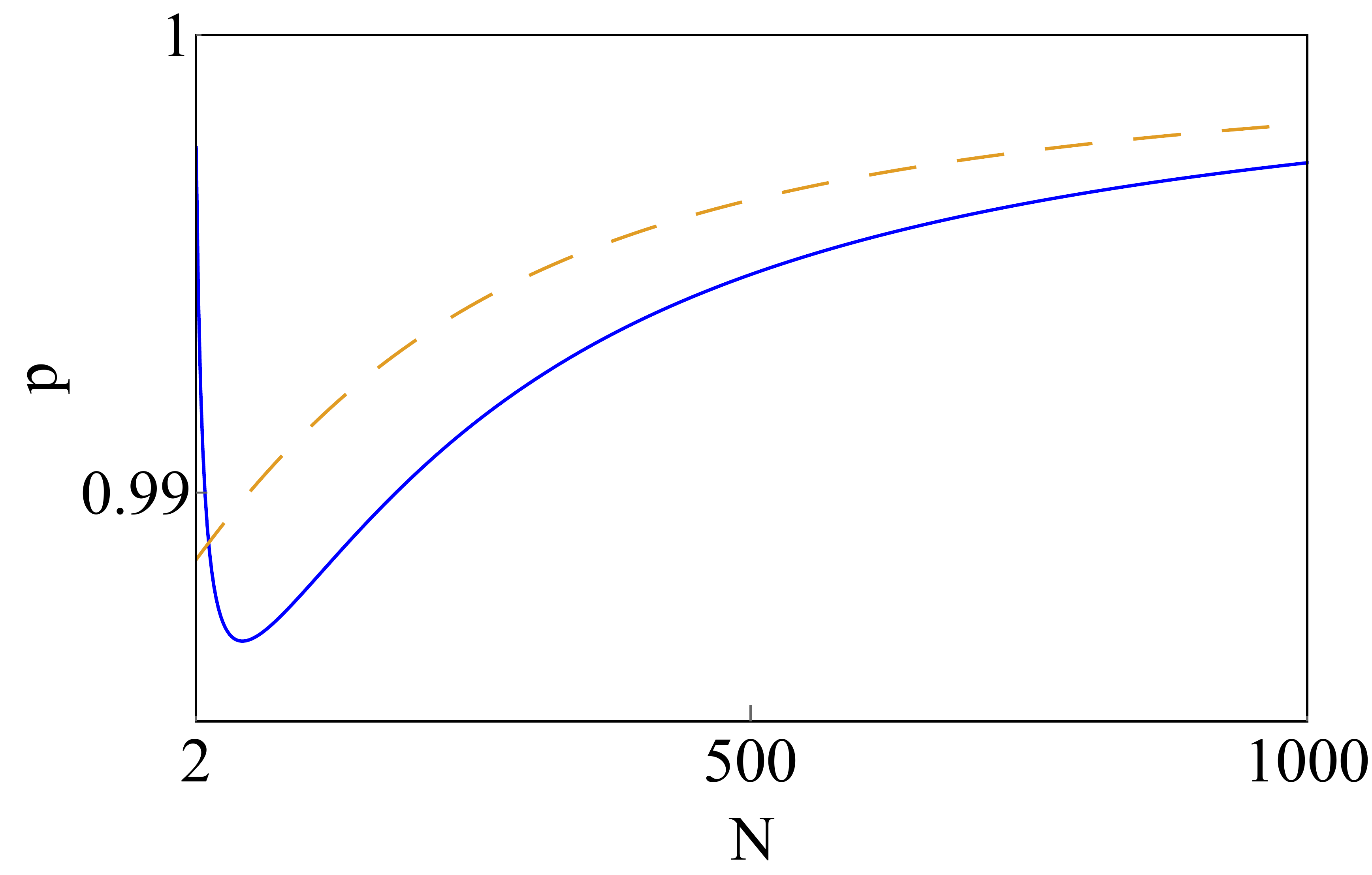}
\caption{Maximum tolerable local white noise for the optimal witnesses given in Eqs. \eqref{S_criterion} (orange dashed line) and \eqref{Duan_criterion} (blue solid line) as a function of the total number of spins. The state that is considered here is a mixture between a spin squeezed state (with a squeezing parameter $\chi t$ given by $10\log_{10}(\xi^2)=-10$ dB for 500 atoms) with probability $p$ and local white noise with probability $1-p$, see Eq. \eqref{noisy_state}. We conclude that the witness \eqref{Duan_criterion} is more resistant to local white noise when $N \geq 30$ for any squeezing between $-1$ and $-10$ dB for 500 atoms and any atom number between 2 and 500.} 
\label{fig3}
\end{figure}

\section{Comparisons of entanglement witnesses using first and second order moments of local collective spin observables}
\label{SectionIV}

The aim of this section is to compare the two optimal witnesses \eqref{S_criterion} and \eqref{Duan_criterion} that we found in the two previous sections. We first compare their resistance with respect to local white noise, then to preparation noise before investigating their resistance to measurement noises.

\subsection{Local white noise}

As a first comparison, we focus on the resistance of the optimal witnesses using first and second order moments of LCSO to local white noise. We compute the maximal amount of noise that can be tolerated by fixing $\chi t=0.0058$ and varying the atom number. The result is shown in Fig. \ref{fig3} where the resistance of the witness \eqref{S_criterion} is drawn in orange (dashed line) and the resistance of the witness \eqref{Duan_criterion} is shown in blue (solid line). Let us recall that smaller $p$ translates into a better resistance to noise. Note also that adding $0.1$\% ($0.5$\%) of local white noise to a spin squeezed state with 500 atoms and $-10$ dB squeezing effectively reduces the squeezing to $-5.6$ dB ($-0.15$ dB). We can fairly say that the witness using second order moments of LCSO has a better resistance to local white noise.\\

While local white noise often corresponds to a worst case scenario, more specific noises are often relevant when one wants to model experiments in detail. In the next section, we compare the two witnesses \eqref{S_criterion} and \eqref{Duan_criterion} with respect to noises that are relevant in experiments using Bose-Einstein condensates.
\subsection{Preparation noise}
To compare the resistance to noise at the preparation level, i.e. before the splitting, we apply the unitary shown in Eq. \eqref{split_state} back into the observables involved in \eqref{S_criterion} and \eqref{Duan_criterion} to get an expression of these witnesses before the splitting. For the witness \eqref{S_criterion}, we get 
\begin{eqnarray}
 \label{cc}
 \nonumber
 S \leq\frac{N(N-1)}{16} \quad \Longleftrightarrow &
 \\
 \frac{\langle (\hat{J}_x^A)^2 \rangle+\langle (\hat{J}_y^A)^2 \rangle -\langle (\hat{J}_z^A)^2 \rangle}{4}-\frac{N}{16}  &\leq\frac{N(N-1)}{16}.
\end{eqnarray}
Here, the expectation values are to be understood on the state before the beam splitter. When considering the subspace that is symmetric under particle interchange, this reduces to 
 \begin{equation} 
 \label{cc}
 S \leq\frac{N(N-1)}{16} \Longleftrightarrow \langle (\hat{J}_z^A)^2 \rangle\geq \frac{N}{4}.
\end{equation}
This shows that any symmetric state having a second moment of a collective spin (in any direction) which is smaller than the one of a coherent spin state with the same mean number of spins leads to entanglement after splitting. Moreover, this entanglement is always detected by the witness \eqref{S_criterion}.

For the witness \eqref{Duan_criterion}, we have 
\begin{equation}
\label{ss}
D\geq 0\Longleftrightarrow \langle(\hat{J}_z^A)^2 \rangle\geq\langle\hat{J}_x^A \rangle-\frac{N}{4}.
\end{equation}
As the maximum value of $\langle\hat{J}_x^A\rangle$ for N spins is $\frac{N}{2}$ \eqref{normeJ}, any state violating \eqref{ss} also violates \eqref{cc}. Therefore the first order witness \eqref{S_criterion} is more robust than the criterion \eqref{Duan_criterion} for any kind of noise before the splitting that keeps the state in the symmetric subspace.

\subsection{Measurement noise: coarse-graining}
As said in the introduction, the local collective observable $\hat{J}_z^A$ is measured by counting the number of atoms in each state 1 and 2, i.e. $\hat{J}_z^A=\frac{\hat{a}_1^\dag \hat{a}_1-\hat{a}_2^\dag \hat{a}_2}{2} = \frac{\hat{n}_1^A-\hat{n}_2^A}{2}$ where $\hat{n}_i^A$ is the atom number at location A in state $i.$ Projections along other spin directions are obtained by appropriate Rabi rotations before the measurement. We here consider the case where the collective spin measurements are coarse-grained due to imperfect atom number measurements. In particular, we assume that the measurement noise leads to an unbiased Gaussian distribution of atom number, i.e. $\hat{n}_i^A$ is replaced by $(\hat{n}_i^A+\epsilon)$ with probability density $g_{\sigma_c}(\epsilon)$, where $\sigma_c^2$ is the variance of the Gaussian noise distribution and similarly for $\hat{n}_i^B.$

Under the assumption that the measurement noise at location A is uncorrelated with the noise in B, the witness involving first order moments of LCSO are insensitive to this noise. Therefore witness \eqref{S_criterion} is insensitive to a coarse-graining of the measurement outcome.

On the contrary, assuming also that the noises on $\hat{n}_1^A$ and $\hat{n}_2^A$ are uncorrelated (similarly in B), the witness involving second order moments of LCSO yields 
\begin{equation}
\langle (\hat{J}_y^A-\hat{J}_y^B)^2   \rangle +\langle (\hat{J}_z^A+\hat{J}_z^B)^2 \rangle  -\langle (\hat{J}_x^A+\hat{J}_x^B)\rangle \geq -2\sigma_{c}^2
\end{equation}
for all separable states. This means for example that for an uncertainty corresponding to 5 atoms $(\sigma_{c}=5),$ a minimum squeezing of $\sim -2$ dB is required to reveal entanglement in a set of 500 atoms with the witness $\eqref{Duan_criterion}.$ 

\subsection{Measurement noise: phase noise} 
\begin{figure}[H] 
\includegraphics[width = 8cm]{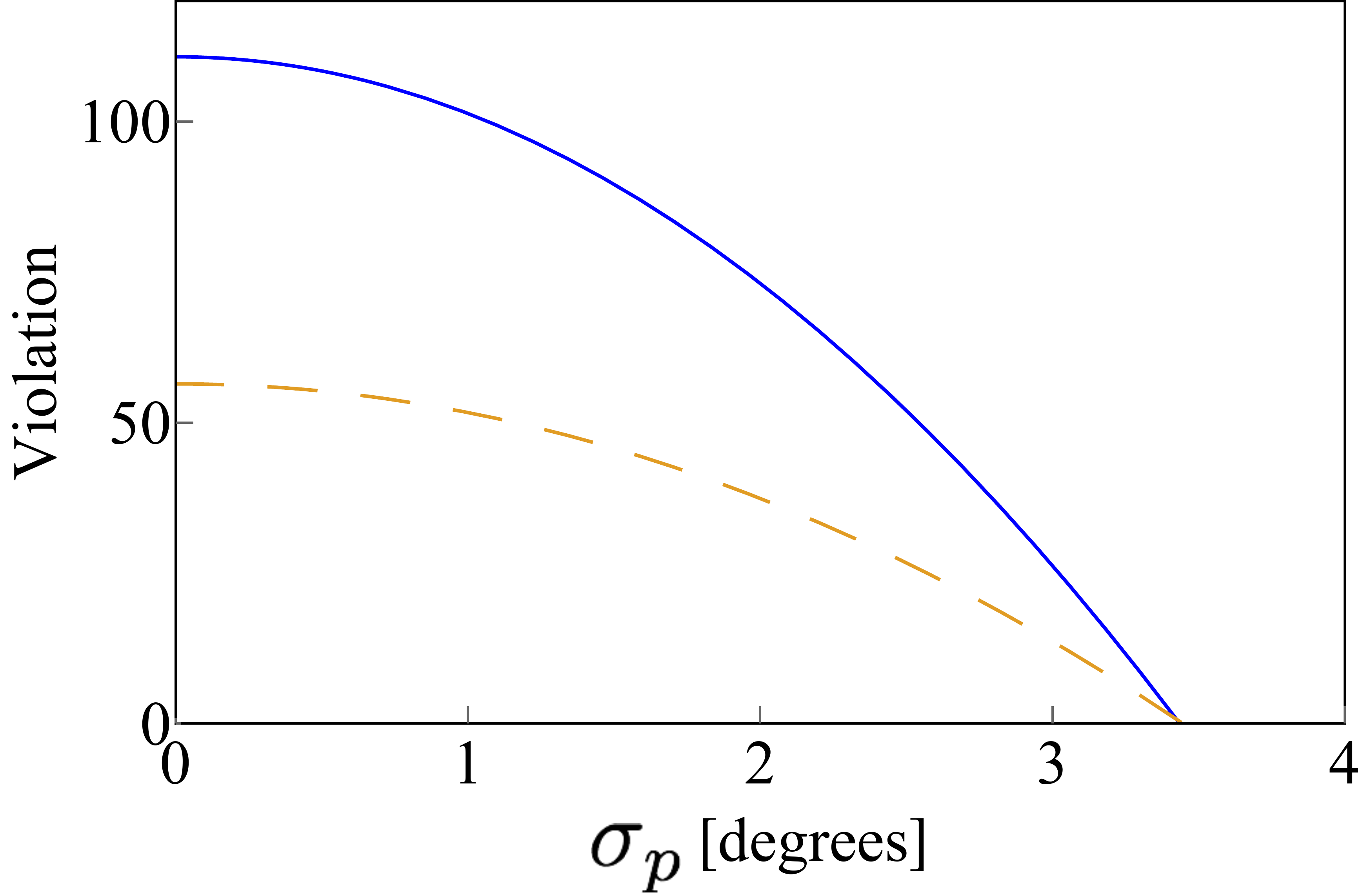}
\caption{The orange dashed line (blue line) gives $S-\frac{N(N+1)}{16}$ ($D$), i.e. the violation of the witness \eqref{S_criterion} (\eqref{Duan_criterion}) as a function of the phase noise (in degrees).}
\label{fig5}
\end{figure}
Due to the difference in energy between the states 1 and 2, the collective spin state $\ket{\psi}$ rotates around the z axis. The spin projections discussed so far are thus implemented in a rotating frame, i.e. the frame of the state is taken as a reference frame. Phase noise refers to a mismatch between the frame of the state and the frame of the measurements which can be due to magnetic field fluctuations. In  the present case, we consider uncorrelated phase noise between the wells. To take this phase noise into account, the spin projections are not calculated on $\ket{\phi}$ but on 
\begin{equation}
\rho_{\sigma}=\int\mathrm{d\theta_A}\mathrm{d\theta_B}g_{\sigma_p}(\theta_A)g_{\sigma_p}(\theta_B)R_A R_B |\phi\rangle \langle \phi | R_A^{-1} R_B^{-1}
\end{equation}
with $R_A=e^{i\theta_A \hat{J}_z^A},$ $R_B=e^{i\theta_B \hat{J}_z^B}$ and $g_{\sigma_p}(\theta_A)$ and $g_{\sigma_p}(\theta_B)$ are unbiased Gaussian distributions with a standard deviation $\sigma_p.$  

Fig. \ref{fig5} shows the violations, i.e. the values of $S-\frac{N(N+1)}{16}$ and $D$ for $-10$ dB squeezing and $N=500$ spins as a function of the standard deviation $\sigma_p$. We see that the witnesses \eqref{S_criterion} and \eqref{Duan_criterion} have essentially the same resistance to phase noise. In particular for phase noise of $\pm 3.4$ degrees, the violation disappears and neither of the witnesses can detect entanglement. We have been able to explore several parameter regimes and for any $\chi t$ between 0.00046258 and 0.0058 which correspond to squeezing between $-1$ and $-10$ dB for 500 atoms and any spin number between 2 and 1000, we found that the violation of both witnesses disappears for the same uncertainties on the phase. We conclude that their resistance to phase noise is thus comparable.


\section{Required statistics}
\label{SectionV}
In this section, we give an estimation of the number of experimental runs that would be necessary to estimate the quantities \eqref{S_criterion} and \eqref{Duan_criterion}. Let us first consider the witness \eqref{S_criterion}. We assume that the spin projections $\hat{J}_i^A \hat{J}_i^B$ are independent quantities that are measured $N_m$ times \cite{discussion}. Let $\bar{X}_k,$ $\bar{Y}_k$ and $\bar{Z}_k$ the values that $\hat{J}_i^A \hat{J}_i^B$ takes at the run $k$ for $i=x,y$ and $z$ respectively. The estimator of $S$ after $N_m$ runs is given by 
\begin{equation}
\bar{S}=\frac{1}{N_m}\sum_{k=1}^{N_m}\bar{X}_k+\frac{1}{N_m}\sum_{i=1}^{N_m}\bar{Y}_k-\frac{1}{N_m}\sum_{k=1}^{N_m}\bar{Z}_k
\end{equation}
and the fluctuations of this mean value are parametrized by 
 \begin{equation}
 \sigma_{\bar{S}}=\frac{1}{\sqrt{N_m}} \sqrt{\sigma_X^2 + \sigma_Y^2 + \sigma_Z^2}
 \end{equation}
where $\sigma_X^2$ is the standard deviation of variables $\bar{X}_k$ and similarly for $\sigma_Y^2$ and $\sigma_Z^2.$ Here we assumed that the runs are independent and identically distributed. Let us consider an experiment performed on the state $\bar \rho.$ The mean value of $\bar S$ after $N_m$ runs is given $\bar S_q=\tr\big(\bar \rho (\hat{J}_x^A\hat{J}_x^B+\hat{J}_y^A\hat{J}_y^B-\hat{J}_z^A\hat{J}_z^B)\big)$ while $\sigma_X^2$ is given by $\sigma_{X,q}^2 = \tr\big(\bar\rho (\hat{J}_x^A\hat{J}_x^B)^2\big)- \Big(\tr \big(\bar\rho \hat{J}_x^A \hat{J}_x^B \big)\Big)^2$ and similarly for $\sigma_Y^2$ and $\sigma_Z^2.$ 
The number of runs that is needed to estimate the value of the witness with a precision 3 times smaller than the distance to the separable bound can thus be estimated by solving
\begin{equation}
|\bar S_q- \frac{N(N+1)}{16}| = \frac{3}{\sqrt{N_m}} \sqrt{\sigma_{X,q}^2 + \sigma_{Y,q}^2 + \sigma_{Z,q}^2}.
\end{equation}
We follow the same line of thought for the criteria D by considering the estimator 
\begin{equation}
\tilde{D}=\frac{1}{N_m}\sum_{k=1}^{N_m}\bar{\bar{X}}_k+\frac{1}{N_m}\sum_{i=1}^{N_m}\bar{\bar{Y}}_k+\frac{1}{N_m}\sum_{k=1}^{N_m}\bar{\bar{Z}}_k
\end{equation}
where $\bar{\bar{X}}_k,$ $\bar{\bar{Y}}_k$ and $\bar{\bar{Z}}_k$ are the values of $-\hat{J}_x^A - \hat{J}_x^B,$ $(\hat{J}_y^A - \hat{J}_y^B)^2$ and  $(\hat{J}_z^A + \hat{J}_z^B)^2$ at the run k.

\begin{figure}[H] 
\includegraphics[width = 8cm]{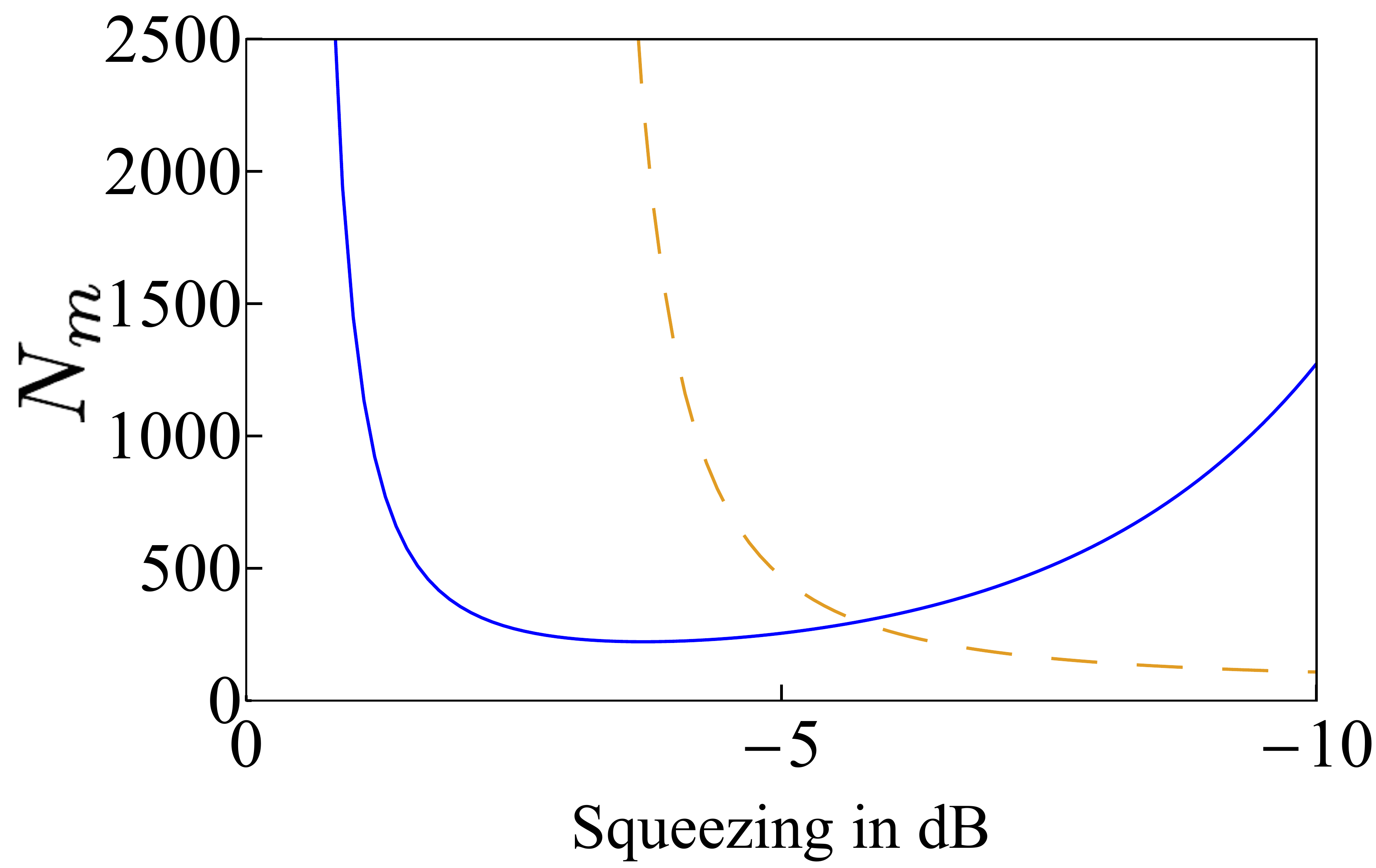}
\caption{Number of evaluations of the witnesses which are required in order to be 3 sigma less than the violation as functions of the initial squeezing in dB. The blue line represents the criterion S and the red dashed line represent the criterion D}
\label{fig6}
\end{figure}

For concreteness, we consider a spin squeezed state made with N=500 spins with an uncertainty on the phase of $\pm 1$ degree and a measurement coarse-graining of $\pm 5$ atoms. As a function of the initial squeezing parameter, we compute the number of runs needed to observe a value of the witnesses \eqref{S_criterion} and \eqref{Duan_criterion} exceeding the separable bound by 3 standard deviations. The result is shown in Fig. \ref{fig6}. We see that one needs less runs to estimate the criteria S with an accuracy of 3 sigma if the initial squeezing $\xi^2 > -6$ dB mostly because of the insensibility with respect to detection noise. \\

\section{Conclusion}
The aim of this work was to clarify the requirements to reveal entanglement between the two parts of a spatially split spin-squeezed Bose-Einstein condensate. We focused on two families of witnesses. The first one uses first order moments of local collective spin operators, i.e.  $\langle\hat{J}_{i}^A\rangle$, $\langle\hat{J}_{i}^B\rangle$,  $\langle\hat{J}_{i}^A \hat{J}_{j}^B\rangle$ where $i, j$ labels the components in the directions $x$, $y$ and $z.$ The second family of witnesses involves not only first order moments of local collective operators, but also the second order moments $\langle(\hat{J}_{i}^A)^2\rangle$ and $\langle(\hat{J}_{i}^B)^2\rangle.$ In both cases, we found the witness that is the most resistant to local white noise. In the first case, we found a witness closely connected to the scalar product given in Ref. \cite{Durkin05}. In the second case, the best linear witness regarding local white noise turns out to be a linear form of the Duan \cite{Duan00} and \cite{Simon00} criteria for spins. We have then compared these two optimal witnesses with respect to their robustness to various noises and we finally gave an estimate of the statistics needed for their experimental measurement. This work lays the theoretical ground that is needed for an ambitious experiment aiming to detect entanglement in a split Bose-Einstein condensate. The next step will be to show how to violate a Bell inequality in this scenario -- a milestone to extend the field of device-independent quantum information processing to many-body physics. 

\section{Acknowledgements} We thank B. Allard, P. Drummond, M. Fadel, A. Peter, M. Reid, P. Sekatski and T. Zibold for valuable discussions and/or comments on the paper. This work was supported by the Swiss National Science Foundation (SNSF), through the NCCR QSIT and the Grant number PP00P2-150579. NS acknowledges the Army Research Laboratory Center for Distributed Quantum Information via the project SciNet.  \\

\bibliographystyle{plain}

\begin{thebibliography}{9}
\bibitem{Amico08} L. Amico, R. Fazio, A. Osterloh, and V. Vedral, Rev. Mod. Phys. {\bf 80}, 517 (2008)

\bibitem{Bloch08} I. Bloch, J. Dalibard, and W. Zwerger, Rev. Mod. Phys. {\bf 80}, 885 (2008)

\bibitem{Mullin08} W. J. Mullin and F. Lalo\"e, Phys. Rev. A 78, 061605(R) (2008)

\bibitem{Laloe09} F. Lalo\"e, and W. J. Mullin, Eur. Phys. J. B 70, 377 (2009)

\bibitem{Gneiting08} C. Gneiting and K. Hornberger, Phys. Rev. Lett. 101, 260503 (2008)

\bibitem{Lewis_Swan15} R.J. Lewis-Swan and K.V. Kheruntsyan, Phys. Rev. A 91, 052114 (2015)

\bibitem{Pelisson16} S. Pelisson, L. Pezz\`e, A. Smerzi, Phys. Rev. A 93, 022115 (2016)

\bibitem{Tura14} J. Tura, R. Augusiak, A.B. Sainz, T. Vertesi, M. Lewenstein, and A. Ac{\'\i}n, Science {\bf 344}, 1256 (2014)

\bibitem{Schmied16} R. Schmied, J-D. Bancal, B. Allard, M. Fadel, V. Scarani, P. Treutlein, and N. Sangouard, Science {\bf 352}, 441 (2016)

\bibitem{Brunner14} N. Brunner, D. Cavalcanti, S. Pironio, V. Scarani, and S. Wehner, Rev. Mod. Phys. {\bf 86}, 419 (2014)

\bibitem{Scarani12} V. Scarani, Acta Physica Slovaca {\bf 62}, 347 (2012)

\bibitem{Ma11} J. Ma, X. Wang, C. P. Sun, and F. Nori, Physics Reports, {\bf 509}, 89 (2011)

\bibitem{Pezze16} L. Pezz\`e, A. Smerzi, M.K. Oberthaler, R. Schmied, and P. Treutlein, arXiv:1609.01609

\bibitem{Kitagawa93} M. Kitagawa and M. Ueda, Phys. Rev. A {\bf 47}, 5138 (1993)

\bibitem{Sorensen01} A. S\o{}rensen, L.-M. Duan, J.I. Cirac, and P. Zoller, Nature {\bf 409}, 63 (2001)

\bibitem{Wineland92} D.J. Wineland, J.J. Bollinger, W.M. Itano, F.L. Moore, and D.J. Heinzen Phys. Rev. A 46, R6797(R) (1992)

\bibitem{Wineland94} D.J. Wineland, J.J. Bollinger, W.M. Itano, and D.J. Heinzen, Phys. Rev. A {\bf 50}, 67 (1994)

\bibitem{Hammerer10} K. Hammerer, A. S\o{}rensen, and E. Polzik, Rev. Mod. Phys. {\bf 82}, 1041 (2010)

\bibitem{He12} Q.Y. He, P.D. Drummond, M.K. Olsen, and M.D. Reid, Phys. Rev. A {\bf 86}, 023626 (2012)

\bibitem{He11} Q.Y. He, M.D. Reid, T. G. Vaughan, C. Gross, M. Oberthaler, and P. D. Drummond, Phys. Rev. Lett. {\bf 106}, 120405 (2011)

\bibitem{Bar2011} N. Bar-Gill, C. Gross, I. Mazets, M. Oberthaler, and G. Kurizki, Phys. Rev. Lett. {\bf 106}, 120404 (2011)

\bibitem{Kurkjian13} H. Kurkjian, K. Paw\l{}owski, A. Sinatra, and P. Treutlein Phys. Rev. A {\bf 88}, 043605 (2013)

\bibitem{Opanchuk12} B. Opanchuk, Q.Y. He, M.D. Reid, and P.D. Drummond Phys. Rev. A {\bf 86}, 023625 (2012)

\bibitem{Duan00} L.-M. Duan, G. Giedke, J.I. Cirac, and P. Zoller, Phys. Rev. Lett. {\bf 84}, 2722 (2000)

\bibitem{Simon00} R. Simon, Phys. Rev. Lett. {\bf 84}, 2726 (2000)

\bibitem{Raymer03} M.G. Raymer, A.C. Funk, B.C. Sanders, and H. de Guise, Phys. Rev. A {\bf 67}, 052104 (2003)

\bibitem{Durkin05} G.A. Durkin and C. Simon, Phys. Rev. Lett. {\bf 95}, 180402 (2005)

\bibitem{Riedel2010} M. F. Riedel, P. B\"ohi,Y. Li, T. W. H\"ansch, A. Sinatra and Philipp Treutlein, Nature {\bf 464}, 1170-1173 (2010)

\bibitem{Gross2010} C. Gross, T. Zibold, E. Nicklas, J. Est\`{e}ve and M. K. Oberthaler, Nature {\bf 464}, 1165-1169 (2010) 

\bibitem{Shin04} Y. Shin, M. Saba, T. A. Pasquini, W. Ketterle, D. E. Pritchard, and A.E. Leanhardt, Phys. Rev. Lett. {\bf 92}, 050405 (2004).

\bibitem{Schumm05} T. Schumm, S. Hofferberth, L.M. Anderson, S. Wildermuth, S. Groth, I. Bar-Joseph, J. Schmiedmayer, and P. Kruger, Nature Phys. {\bf 1}, 57 (2005)

\bibitem{Reinaudi2007} G. Reinaudi, T. Lahaye, Z. Wang, and D. Guery-Odelin, Opt. Lett. {\bf 32}, 3143 (2007)

\bibitem{Julsgaard01} B. Julsgaard, A. Kozhekin, and E.S. Polzik, Nature {\bf 413}, 400 (2001)

\bibitem{discussion} We emphasize that $N_m$ is the number of runs assuming that $\hat{J}_x^A \hat{J}_x^B,$ $\hat{J}_y^A \hat{J}_y^B,$ and $\hat{J}_z^A \hat{J}_z^B$ are measured at each run. In the case where these quantities are measured separately, $3N_m$ gives a conservative estimation of the statistics needed to estimate the considered witnesses.

\end{thebibliography}

\end{document}